\documentclass[journal]{IEEEtran}
\usepackage{cite}
\usepackage{multirow}
\usepackage{diagbox}

\ifCLASSINFOpdf
  \usepackage[pdftex]{graphicx}
  \usepackage[caption=false]{subfig}
  \DeclareGraphicsExtensions{.pdf,.jpeg,.png}
\else
\fi
\newcommand{\statsig}{Horizontal bars link systems for which $\textit{p-value} > 0.05$.}

\IEEEoverridecommandlockouts

\usepackage{tikz}
\usepackage{textcomp}
\usepackage{hyperref}
\usepackage{lipsum}

\newcommand\copyrighttext{%
  \footnotesize \textcopyright 2019 IEEE.  Personal use of this material is permitted.  Permission from IEEE must be obtained for all other uses, in any current or future media, including reprinting/republishing this material for advertising or promotional purposes, creating new collective works, for resale or redistribution to servers or lists, or reuse of any copyrighted component of this work in other works. DOI: \href{https://doi.org/10.1109/LSP.2019.2961213}{10.1109/LSP.2019.2961213}}
\newcommand\copyrightnotice{%
\begin{tikzpicture}[remember picture,overlay]
\node[anchor=south,yshift=5pt] at (current page.south) {\fbox{\parbox{\dimexpr\textwidth-\fboxsep-\fboxrule\relax}{\copyrighttext}}};
\end{tikzpicture}%
}

\begin{document}
%
\title{Voice Conversion for Whispered Speech Synthesis}
%
%
%

\author{Marius~Cotescu,
			Thomas~Drugman,~\IEEEmembership{Member,~IEEE,}
			Goeric~Huybrechts,
			Jaime~Lorenzo-Trueba,~\IEEEmembership{Member,~IEEE,}
			and~Alexis~Moinet%
\thanks{M. Cotescu, T. Drugman, G. Huybrechts, J. Lorenzo-Trueba and A. Moinet are with Amazon.com (e-mail: cotescu@amazon.com)}
\thanks{Manuscript received December 11, 2019.}}

%
%

\markboth{IEEE Signal Processing Letters}%
{Shell \MakeLowercase{\textit{et al.}}: Bare Demo of IEEEtran.cls for IEEE Journals}
%



\maketitle
\copyrightnotice
\begin{abstract}
We present an approach to synthesize whisper by applying a handcrafted signal processing recipe and Voice Conversion (VC) techniques to convert normally phonated speech to whispered speech. We investigate using Gaussian Mixture Models (GMM) and Deep Neural Networks (DNN) to model the mapping between acoustic features of normal speech and those of whispered speech. We evaluate naturalness and speaker similarity of the converted whisper on an internal corpus and on the publicly available wTIMIT corpus. We show that applying VC techniques is significantly better than using rule-based signal processing methods and it achieves results that are indistinguishable from copy-synthesis of natural whisper recordings. We investigate the ability of the DNN model to generalize on unseen speakers, when trained with data from multiple speakers. We show that excluding the target speaker from the training set has little or no impact on the perceived naturalness and speaker similarity of the converted whisper. The proposed DNN method is used in the newly released Whisper Mode of Amazon Alexa.\looseness=-1
\end{abstract}

\begin{IEEEkeywords}
whispered speech conversion, voice conversion (VC), whispered text to speech (TTS).
\end{IEEEkeywords}

\vspace{-1mm}
\section{Introduction}

\IEEEPARstart{W}{hisper}  is a natural way of speaking, primarily characterized by the lack of phonation \cite{Monoson1984}.  As Text-to-Speech (TTS) systems and virtual assistants become more natural and widespread, the scenarios when users need to interact with the devices in a quiet or private manner increase. In these cases, a user may wish to whisper to the device and would expect a response in a whispered voice, as is the case of the recently released Whisper Mode of Amazon Alexa \cite{whispermode}. While the generation of normally phonated speech from text (i.e. TTS), from the speech of other speakers (i.e. voice conversion), or even from whispered speech (e.g. pathological applications) is well researched, there has been little effort in developing technologies that can produce whispered speech.\looseness=-1

In this paper, we aim at converting normal speech into whispered speech while maintaining high naturalness and speaker identity. To the best of our knowledge, this is the first study attempting to achieve this goal. For this purpose, two main approaches are investigated. First, we create a hand-crafted digital signal processing (DSP) recipe that simulates acoustic modifications in a source-filter based vocoder \cite{fant1970acoustic, morise2016world}. Second, we learn the mapping between normal and whispered acoustic features using voice conversion (VC) techniques. \looseness=-1

Physiologically speaking, whispered speech differs from normal speech mainly by the lack of vocal folds vibration. Exhaled air passes directly through the restricted but open larynx which causes a turbulent aperiodic glottal airflow \cite{Monoson1984}. The acoustic manifestations of whispered speech are however more diverse than a simple lack of pitch and harmonics. 
Extensive studies of the acoustic characteristics of whispered speech \cite{kallail1984formant,sharifzadeh2012comprehensive,jovivcic2008acoustic,itoh2001acoustic,tartter1989s} found consistent changes in the formant structure, spectral tilt and consonant energy in whispered speech when compared to normal speech. Some of these findings were used to develop whispered speech detectors \cite{raeesy2018lstm, zhang2009advancements}. We also draw on these findings to build a rule based DSP recipe that transforms normal speech into whispered speech.\looseness=-1

VC models have traditionally relied on GMMs \cite{stylianou1998continuous, toda2007voice, helander2010voice} to learn the joint distribution of acoustic features for source and target speakers from parallel corpora. More recently, feed forward, recurrent, and convolutional neural networks approaches \cite{mohammadi2014voice,saito2017voice,sun2015voice,tanaka2018atts2s}  take advantage of larger parallel speech corpora to produce high-quality conversions. Other techniques achieve voice conversion by learning common latent \cite{Hsu2017, kameoka2018acvae, huang2018refined,fang2018high, kameoka2018stargan} or explicit distributions \cite{Tian2018} from non-parallel corpora. The task of converting whispered speech to normal speech has received significant research interest. There are many examples of applying VC techniques to the problem that vary from code excited linear prediction to recurrent neural networks \cite{sharifzadeh2010reconstruction,Meenakshi2018} and we draw inspiration from them.\looseness=-1

The main contributions of this paper are: 
\emph{i)} we explore the problem of normal to whispered speech conversion; 
\emph{ii)} we propose a rule-based DSP system; 
\emph{iii)} we use GMM and DNN models as VC techniques to learn the more complex mapping functions; 
\emph{iv)} we evaluate our proposed solutions both on an internal proprietary corpus and on the publicly available wTIMIT\footnote{The wTIMIT corpus is publicly available and can be downloaded from http://www.isle.illinois.edu/sst/data/wTIMIT/.} corpus \cite{lim2011computational}; 
\emph{v)} we show that VC models can produce converted whisper of the same quality as vocoded natural recordings of whisper; 
\emph{vi)} we show that through multi-speaker training, DNN models have the ability to generalize to unseen speakers.\looseness=-1

\vspace{-3mm}
\section{DSP-based Conversion System}
\label{sec:dsp}
We implemented a rule-based DSP transformation that applies the main acoustic modifications of whispered speech in the paradigm of the source-filter model. As vocoder, we opted for WORLD \cite{morise2016world} which is widely used in speech synthesis for its high quality and computational efficiency. 
In addition to the obvious unvoiced setting, three main spectral changes were applied: \emph{1)} removing the glottal shaping, \emph{2)} shifting of the first formant, \emph{3)} increasing the formant bandwidth. 
\emph{Step 1} removes the spectral coloring introduced by the voiced glottal air flow  \cite{doval2006spectrum, drugman2014glottal} by subtracting the typical spectrum of a glottal pulse based on the LF model \cite{fant1985four} from the spectral envelope. 
In \emph{Step 2} we follow the findings reported in \cite{sharifzadeh2012comprehensive} and we use frequency warping of the spectral envelope to shift the first formant upwards by about 100 Hz while keeping the second and third formants unchanged.
In \emph{Step 3} we apply a moving average filtering to the log spectrum with a 400 Hz-wide triangular window to mimic the increased formant bandwidth reported in \cite{tartter1989s, sharifzadeh2012comprehensive}.\looseness=-2

Although the DSP-based system may look simple, it can generalize on a diversity of male, female and child speakers. It was released in 2017 to support the Amazon Polly 'whispered' SSML tag \cite{whisperedssml} and is applied to 59 voices in 29 languages.

\vspace{-2mm}
\section{Voice Conversion System}
\label{sec:whisper_conversion}
We propose a conversion system based on the architecture presented in \cite{Kobayashi2018}. We extend it with the ability to use a feed-forward DNN conversion model and to train on multiple source and target speakers. A similar method was also reported in \cite{NishaMeenakshi2018ReconstructionOA}. Our approach differs in that we apply it to acoustic data of multiple speakers to convert normal speech to whispered speech.\looseness=-2

We start with one parallel corpus containing normal speech and whispered recordings of the same texts. Each pair of files is analyzed and acoustic features are extracted. The acoustic features frames are aligned using Dynamic Time Warping (DTW) in the iterative fashion described in \cite{Kobayashi2018}. 
The aligned features are used to train either a DNN or a GMM conversion model.
At inference time, the source acoustic features are extracted from the input waveform and transformed by the conversion model. The converted features are then synthesized with the WORLD \cite{morise2016world} vocoder.\looseness=-2

\vspace{-5mm}
\subsection{Feature Extraction}
Some VC methods choose to have separate feature sets and models for the source and vocal tract features. We choose to use a single set of features because \emph{i)} whispered speech does not have pitch and \emph{ii)} the voiced source characteristics inform the spectral envelope transformations. We use 80 mel-frequency cesptral coefficients (MFCC) \cite{imai1983cepstral}. 
The cepstral domain provides a compact and untangled representation of both the vocal tract and the excitation, where the lower order coefficients are related to the vocal tract shape and the higher order coefficients describe excitation source traits (e.g. voicing strength, pitch) \cite{oppenheim1968homomorphic}. The cepstral coefficients are computed on 50 ms Hanning analysis windows, with a frame shift of 5 ms. All recordings were down-sampled to 24 kHz. \looseness=-2

\vspace{-5mm}
\subsection{GMM Whisper Conversion}
 
We used \textit{sprocket} \cite{Kobayashi2018}, an open-source implementation of GMM-based voice conversion \cite{toda2007voice}. It was used as a baseline in the 2018 Voice Conversion Challenge \cite{Lorenzo-Trueba2018} and it thus gives us a solid anchor when comparing to other VC methods. We tuned the hyper parameters of the GMM conversion model on the internal validation set (see Section \ref{ssec:datasets}). We observed the best results using a 64-mixture GMM with the convergence threshold set at $0.001$. At inference time, we used Maximum Likelihood Parameter Generation (MLPG) considering the Global Variance (GV) \cite{toda2007speech} to generate the converted target cepstral coefficients.\looseness=-2

\vspace{-5mm}
\subsection{DNN Whisper Conversion}

In this paper, we propose using a DNN model to learn a general mapping from acoustic features of normal speech to those of whispered speech. Neural networks can be trained on batches \cite{bottou1991stochastic} and benefit from several regularization techniques \cite{krogh1992simple,srivastava2014dropout,han2015learning} that allow them to generalize better on large diverse training sets.\looseness=-2

We use a neural network with 4 hidden layers, of size (128, 64, 64, 128) with a rectified linear activation function. We use z-score normalization for both the input and output feature vectors. At inference time, the estimated target vector is scaled back, before being sent to the processing chain. We train the network using the Adam \cite{kingma2014adam} optimizer to minimize the L2 loss. We use a batch size of 2048 frames, a learning rate of 0.002, and L2 weights regularization \cite{krogh1992simple}, with $\lambda = 10^{-5}$. The hyper parameters were tuned on the internal validation set, described in Section \ref{ssec:datasets}. \looseness=-2

\vspace{-3mm}
\section{Experimental Protocol}
\label{sec:experimental_protocol}
\vspace{-3mm}
\subsection{Datasets}
\label{ssec:datasets}
\vspace{-1mm}

We used two corpora to train and evaluate our systems. The first one is an internal corpus of parallel whispered and normal speech recordings. It contains recordings of phonetically rich texts from 5 English female speakers, each from a different locale: Australia (AU), Canada (CA), Great Britain (GB), India (IN) and United States (US). All speakers are professional voice talents and the recordings were done with high quality equipment. The corpus contains 10 hours of both normal and whispered speech (AU 2h, CA 3h, GB 1.5h, IN 3h, US 0.5h). \looseness=-2

The second dataset is a subset of the wTIMIT \cite{lim2011computational} corpus. It contains parallel recordings of normal and whispered speech from 50 male and female English speakers from Singapore (SG) and the United States (US). The recordings average at 50 minutes per speaker. The SNR and quality of the recordings varies greatly between speakers. To address this problem, we manually selected 25 speakers for which the recordings are decently clean. The clean subset contains 20 US speakers and 5 SG speakers (13 male and 12 female). \looseness=-2

For both datasets, we randomly split the data of each speaker into a training, a validation and a test set. We used a 80/10/10\% split, respectively.

\vspace{-4mm}
\subsection{Systems}
\vspace{-1mm}

Table \ref{tab:systems} shows the technologies and the data used to train the different conversion models. The \textbf{Oracle} system consists of copy-synthesis samples reconstructed with the WORLD \cite{morise2016world} from cepstral coefficients computed on original whisper recordings. It constitutes a strong baseline to which we report all our proposed methods. We trained speaker-dependent DNN (\textbf{SD})  and GMM (\textbf{sprocket}) models for each speaker in the two datasets. The \textbf{All} models were trained on all the speakers from a given dataset (i.e. either wTIMIT or the internal one), while the \textbf{Excl} models were trained on all the speakers in the dataset, except the target speaker. The \textbf{DSP} samples are synthesized using the technique described in Section \ref{sec:dsp}. \looseness=-2

\begin{table}[t]
  \caption{Technology and data used for the evaluated systems}
  \label{tab:systems}
  \centering
  \begin{tabular}{p{0.12\columnwidth}p{0.12\columnwidth}p{0.56\columnwidth}}
    \hline
          \textbf{Name} & \textbf{Tech.} & \textbf{Data}     \\
    \hline
    	\textbf{Rec} & - & Natural whisper recordings \\
    	\textbf{Oracle} & Vocoder & Vocoded whisper recordings \\
    	\textbf{sprocket} & GMM & Only the evaluated speaker  \\
    	\textbf{SD} & DNN & Only the evaluated speaker \\
		\textbf{All} & DNN & The entire internal or wTIMIT corpus\\
		\textbf{Excl} & DNN & All speakers in the internal or wTIMIT corpus, excluding the evaluated speaker  \\
		\textbf{DSP} & DSP & - \\
    \hline
  \end{tabular}
\end{table}

\vspace{-4mm}
\subsection{Listening Tests}
\vspace{-1mm}

We used naturalness listening tests to measure the general quality of the conversion and speaker similarity tests to confirm that the speaker identity is maintained. We used self-reported effort to measure the intelligibility of the proposed systems. We used modified MUltiple Stimuli with Hidden Reference and Anchor (MUSHRA) \cite{itu20031534} for the naturalness and similarity tests and a Mean Opinion Score (MOS) for the intelligibility test. We modified the MUSHRA tests by not enforcing that at least one system gets the maximum rating. \looseness=-2

For the naturalness tests, we asked the listeners to evaluate the stimuli in terms of their naturalness on a scale from 0 -- "Completely unnatural" to 100 -- "Completely natural".
In the speaker similarity tests, in addition to the evaluated stimuli, we also presented the listeners with a reference recording of the target speaker whispering a different phrase. We asked the listeners to score the speaker similarity of the stimuli to the provided reference on a scale from  0 -- "A completely different speaker" to 100 -- "The same speaker".
In the intelligibility test we asked the listeners to evaluate the effort they needed to understand the message. They reported on a scale from 1 "Very difficult: I cannot understand what was said" to 7 "Very easy:  I understood everything". We suggested 4 as the middle point "Fair:  I understood everything, but I had to focus". \looseness=-2

For each speaker we randomly chose 50 phrases from the test set, which were used in all the listening tests.
We recruited listeners on the Amazon Mechanical Turk platform and each listener evaluated exactly 10 phrases. We had no guarantee of their English proficiency, or the equipment they used. 
Following the guidelines in \cite{koo2016guideline} we use Intra-Class Correlation (ICC(2,k)) \cite{donner1980estimation} to measure the reliability of our results. The test reliability is said to be good if the ICC value is above 0.75 and excellent if greater than 0.90 \cite{koo2016guideline}. We present the value in all boxplots. 
 We use paired Student t-tests with Holm-Bonferroni correction to validate the statistical significance of the differences between systems, considering it validated when $\textit{p-value} < 0.05$. \looseness=-2

\vspace{-3mm}
\section{Results}
\label{sec:results}

\vspace{-1mm}
\subsection{General Quality}
\label{ssec:general_quality}
\vspace{-1mm}

\begin{figure}[t]
  \centering
  \includegraphics[width=0.95\linewidth]{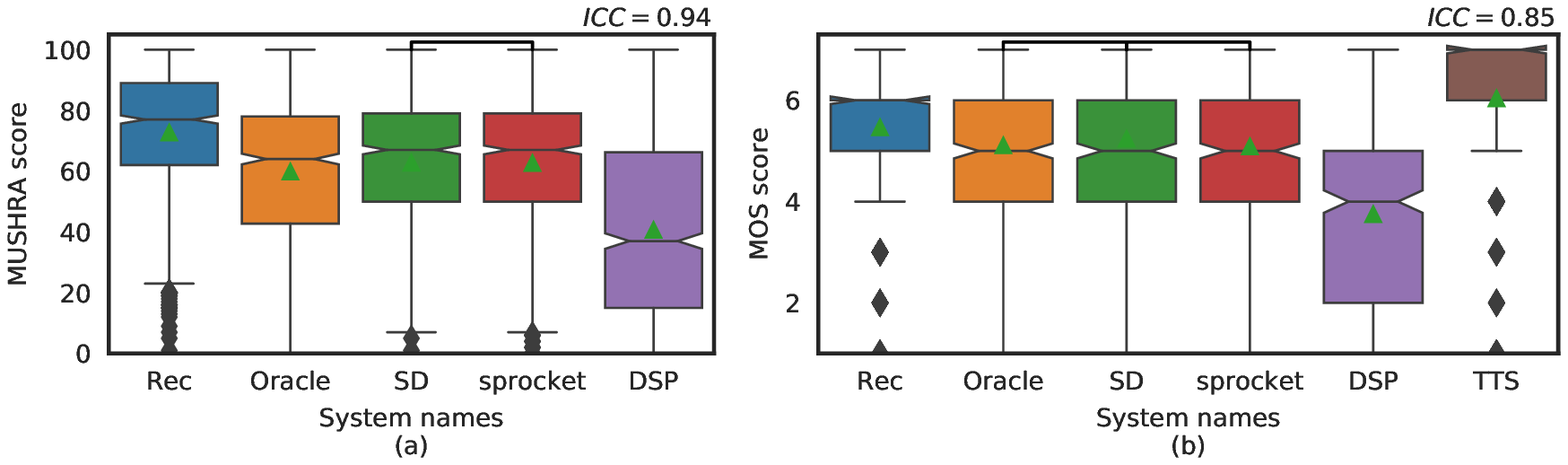}
  \vspace{-2mm}
  \caption{Naturalness (a) and intelligibility (b) scores for the US voice. \statsig}
  \label{fig:general_quality}
\end{figure}

We assessed the general quality of the proposed techniques by comparing them to the copy-synthesis system \textbf{Oracle}. We conducted naturalness MUSHRA and intelligibility MOS listening tests for the US internal speaker. We asked 100 listeners to evaluate  samples from five systems: \textbf{Rec}, \textbf{Oracle}, \textbf{SD}, \textbf{sprocket}, and \textbf{DSP}.  
In the intelligibility test we included samples of normal speech synthesized using the Amazon Polly service (\textbf{TTS}) as a high anchor.
Fig.~\ref{fig:general_quality} shows the boxplots for the two listening tests. \looseness=-2

Both VC methods perform much better than the handcrafted DSP method and achieve the maximum expected naturalness and intelligibility. They performed better or equal to the Oracle system, reaching the limits of the technology. In single speaker scenarios one can use any of the two methods with equal results. \looseness=-2

Listeners find it harder to understand whispered speech than they do normal speech \cite{freyman2012intelligibility}. This is demonstrated by the big difference between the \textbf{TTS} system and the whispered recordings. However, both VC methods are consistently scoring above 4. \looseness=-1

\vspace{-4mm}
\subsection{Multi-Speaker Conversion}
\label{ssec:internal_results}
\vspace{-1mm}

In this section we evaluate the ability of the DNN model to generalize to unseen speakers when trained on data from multiple speakers. Although both the DNN and the GMM systems performed similarly in the speaker independent task, we could not train GMM models on multi-speaker datasets. Multiple speakers datasets contained unpredictable modalities that the GMM model could not capture well. We tried to apply \textbf{SD} models to unseen speakers and had unstable results so we will not present evaluations of this scenario either. \looseness=-2

We conducted naturalness and speaker similarity MUSHRA tests on both the internal corpus and the wTIMIT corpus. We observed the same quality patterns for all wTIMIT speakers in informal listening tests, independent of the recording conditions or accent. However, for the formal listening tests we chose two male (M107, M111) and two female (F116, F120) speakers with clean recording environments for both normal and whispered sets. \looseness=-2

\subsubsection*{Intra-Dataset Generalization}

\begin{figure}[t]
  \centering
  \includegraphics[width=0.95\linewidth]{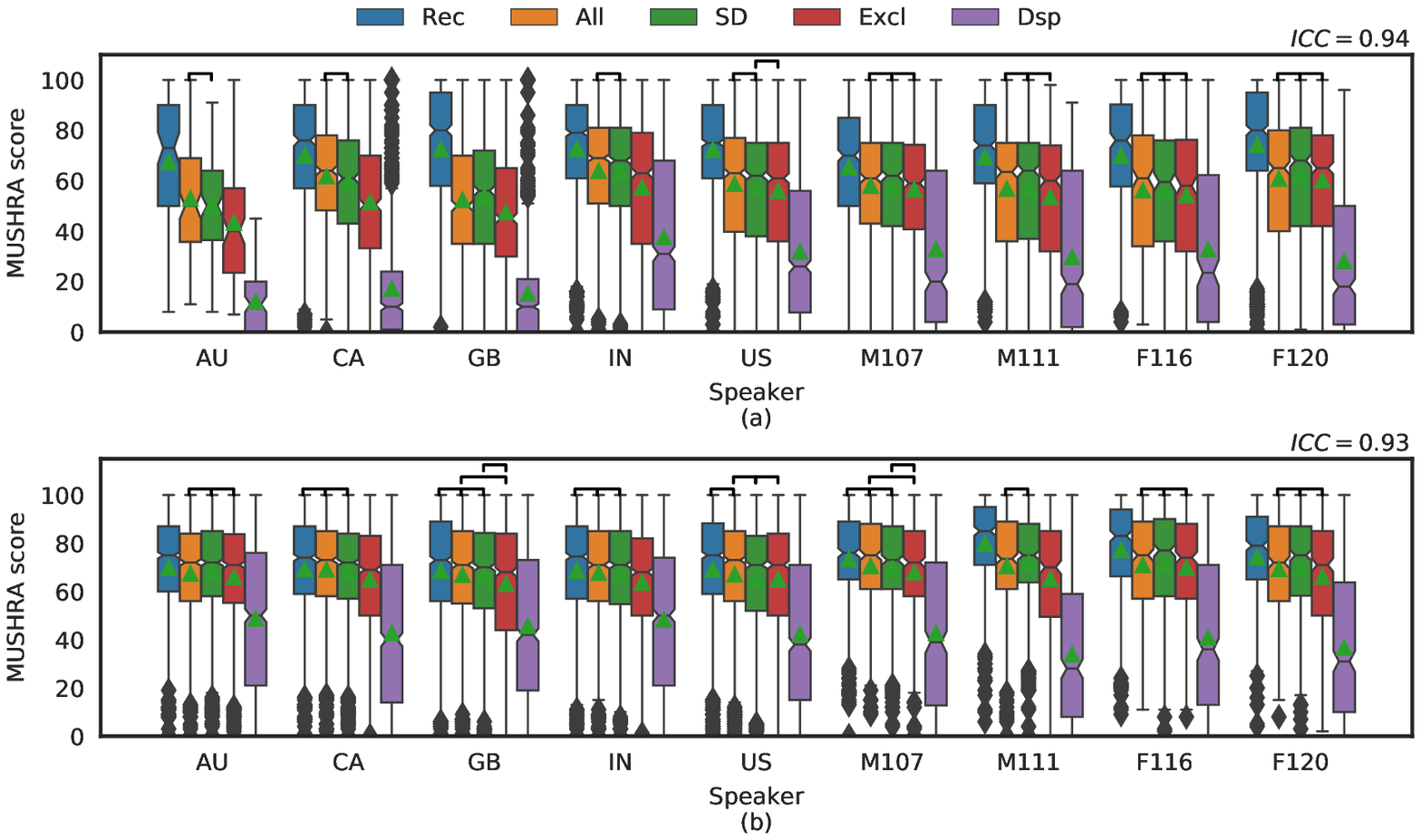}
  \vspace{-2mm}
  \caption{Intra-dataset naturalness (a) and speaker similarity (b) scores. \statsig}
  \label{fig:generalization}
\end{figure}

We evaluated samples from five systems: \textbf{Rec}, \textbf{All}, \textbf{SD}, \textbf{Excl} and \textbf{DSP}. The \textbf{All} and \textbf{Excl} models were trained only on speakers from the same dataset as the evaluated speaker. We used 100 listeners for each test on internal speakers and 40 listeners for the wTIMIT speakers. Fig.~\ref{fig:generalization} shows the boxplots of the MUSHRA scores in the naturalness and speaker similarity tests for the internal and wTIMIT speakers.\looseness=-2

The ranking of the systems' average naturalness and speaker similarity scores is consistent across most of the speakers in both datasets. The reference system \textbf{Rec} is getting the highest scores, followed by a close grouping of the multi-speaker (\textbf{All}) and the speaker dependent (\textbf{SD}) systems. The differences in naturalness and speaker similarity between the two models are not statistically significant, except in the case of the internal GB speaker.  The multi-speaker (\textbf{Excl}) systems excluding the target speaker follow closely. We observe that removing the target speaker from the training set has little effect in the internal corpus and almost no effect in the wTIMIT corpus. The naturalness and speaker similarity scores of the \textbf{Excl} system are slightly lower than those of  \textbf{All}. There is no statistically significant difference in naturalness for five speakers and in speaker similarity for six speakers. The \textbf{DSP} system has the lowest mean score in all tests. \looseness=-2

Training a model with data from more speakers does not lead to a significant increase in quality on seen speakers when compared to speaker dependent models. This is consistent with the results from Section \ref{ssec:general_quality}.  However, it has a positive impact on the model's ability to generalize to unseen speakers. The \textbf{Excl} models trained on internal speakers can produce whisper for unseen speakers with only a small drop in quality, independent of their accent, while the ones trained on the wTIMIT corpus are performing as well as the \textbf{All} and the \textbf{SD} models. We believe that the more varied recording conditions and speaker characteristics present in the wTIMIT corpus helps the models learn a more general transformation.\looseness=-2

\subsubsection*{Cross-Dataset Generalization}

\begin{figure}[t]
  \centering
  \includegraphics[width=\linewidth]{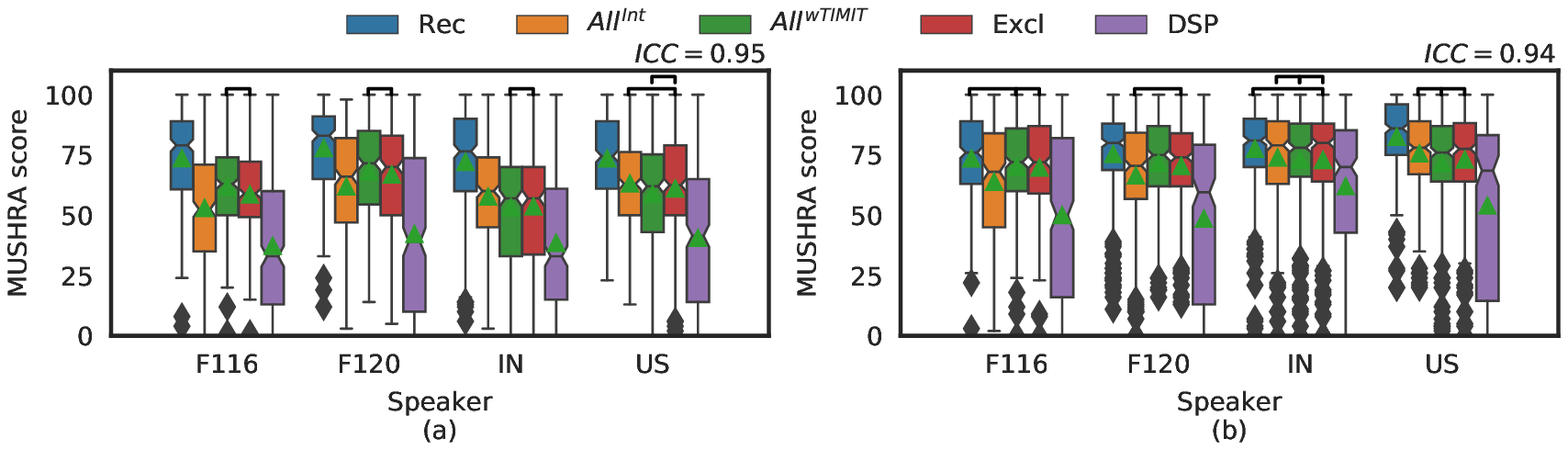}
	\vspace{-6mm}
  \caption{Cross-dataset naturalness (a) and speaker similarity (b) scores. \statsig}
  \label{fig:cross_set}
  \vspace{-1mm}
\end{figure}

We assessed the ability of the systems to generalize across the two datasets. We asked 30 listeners to evaluate the naturalness and speaker similarity of five systems: \textbf{Rec}, $\mathbf{All^{Int}}$, $\mathbf{All^{wTIMIT}}$, \textbf{Excl} and \textbf{DSP} systems. The \textbf{Excl} system was always trained on the dataset of the target speaker. Because there are no male speakers in the internal corpus, we only evaluated the two wTIMIT female speakers (F116, F120). We chose the IN and US internal speakers for formal evaluation as they performed the best in the intra-dataset evaluation. Informal listening tests confirmed the same quality pattern across all internal speakers. Fig.~\ref{fig:cross_set} shows the boxplots of the scores.\looseness=-2

We find that the models need diverse data in order to generalize to more diverse conditions. Although the $\mathbf{All^{Int}}$ model is performing much better than the handcrafted DSP system on the wTIMIT speakers, it is still significantly worse than the \textbf{Excl} system trained on wTIMIT data. This is caused partly by the fact that the internal corpus only contains 5 speakers and partly by the fact that the wTIMIT recordings are noisier. The noise in the wTIMIT corpus also prevented the $\mathbf{All^{wTIMIT}}$ model from performing better on the internal corpus. However, the presence of more varied data helped it generalize even to unseen recording conditions and it performed on par with the \textbf{Excl} models on the internal speakers. \looseness=-2

\subsubsection*{Cross-Gender Generalization}

\begin{figure}[t]
  \centering
  \includegraphics[width=\linewidth]{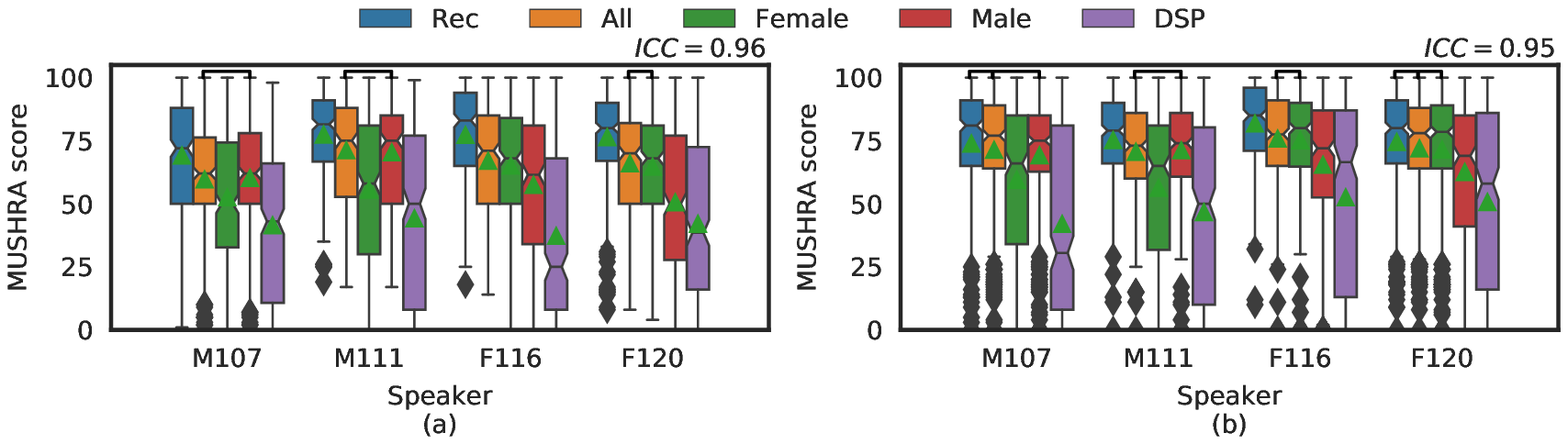}
  \vspace{-6mm}
  \caption{Cross-gender naturalness (a) and speaker similarity (b) scores. \statsig}
  \label{fig:cross_gender}
  \vspace{-1mm}
\end{figure}

We assessed the ability of the DNN models to generalize across genders in the wTIMIT corpus. We asked 30 listeners to evaluate the naturalness and speaker similarity of five systems: \textbf{Rec}, \textbf{All}, \textbf{Female}, \textbf{Male} and \textbf{DSP} systems for the four wTIMIT evaluation speakers. The \textbf{All} model was trained on the entire wTIMIT corpus, while the \textbf{Male} and \textbf{Female} models were trained on all the male and female wTIMIT speakers, respectively. Fig.~\ref{fig:cross_gender} shows the boxplots of the MUSHRA scores. \looseness=-2

The gender balance is less important when the target speaker is present in the training set: the matched gender models perform as well as the \textbf{All} model. However, the model seemed to need gender balanced data in order to generalize well to unseen speakers from both genders. Using models trained on either male of female speakers on unseen speakers of the other gender produced whisper in every case. However, the gender identity of the converted samples was affected and the quality was varying, as reflected by the lower means and higher variances of the naturalness and speaker similarity scores. This drop cannot be explained by the data reduction, as the cross-dataset generalization experiments show less degradation when applying the model trained on the internal corpus to the wTIMIT female speakers.  \looseness=-2

The poor performance in the cross-gender scenario is caused by the fact that the acoustic characteristics of the voiced excitation in the cepstral space vary significantly between the genders. They vary not only in pattern, but also in where they occur in the cepstrum (i.e. the inverse relationship between pitch and quefrency). We highly recommend using gender balanced training datasets in applications where cross gender generalization is important.\looseness=-2

\vspace{-3mm}
\section{Conclusions}
\label{sec:conclusions}
\vspace{-1mm}
This paper was, to the best of our knowledge, the first study attempting to convert normal speech into whispered speech. We investigated using signal processing and two VC techniques (based on GMM and DNN  models) to achieve this goal.  We evaluated the three methods on both an internal corpus and on the wTIMIT public corpus. \looseness=-2

We found out that the DSP recipe had a robust performance over all speakers from both corpora, however it is far from the theoretical limit imposed by the vocoder. All VC approaches however outperformed the DSP system in naturalness, intelligibility and speaker similarity evaluations. They reached the technical limit imposed by the vocoder and feature extraction chain. We showed that the DNN models can learn a speaker-independent mapping when trained on multiple speakers and that they are able to generalize and produce whispered speech for unseen speakers. We found that the DNN models can be robust to recording conditions if trained with varied enough data. The DNN model cannot generalize across genders and a gender balanced corpus is recommended for cross gender applications. The proposed DNN method was integrated into Amazon Alexa, and is used to generate the output for the newly released Whisper Mode \cite{whispermode}.\looseness=-2

\bibliographystyle{IEEEtran}

\bibliography{references}

\end{document}